An Integrated Constrained Gradient Descent (iCGD) Protocol to Correct Scan-Positional Errors for Electron Ptychography with High Accuracy and Precision

Ning Shoucong[1], Wenhui Xu [2,3], Leyi Loh[1], Zhen Lu[4], Michel Bosman[1], Fucai Zhang[2], Qian He[1]*

1. Department of Materials Science and Engineering, National University of Singapore, 117575, Singapore
2. Department of Electrical and Electronic Engineering, Southern University of Science and Technology, Shenzhen 518055, China
3. Harbin Institute of Technology, Harbin, 150001, China
4. Institute of Physics, Chinese Academy of Sciences, Beijing 100190, China
   Email: heqian@nus.edu.sg

**Abstract:**

Correcting scan-positional errors is critical in achieving electron ptychography with both high resolution and high precision. This is a demanding and challenging task due to the sheer number of parameters that need to be optimized. For atomic-resolution ptychographic reconstructions, we found classical refining methods for scan positions not satisfactory due to the inherent entanglement between the object and scan positions, which can produce systematic errors in the results. Here, we propose a new protocol consisting of a series of constrained gradient descent (CGD) methods to achieve better recovery of scan positions. The central idea of these CGD methods is to utilize *a priori* knowledge about the nature of STEM experiments and add necessary constraints to isolate different types of scan positional errors during the iterative reconstruction process. Each constraint will be introduced with the help of simulated 4D-STEM datasets with known positional errors. Then the integrated constrained gradient decent (iCGD) protocol will be demonstrated using an experimental 4D-STEM dataset of the $1H\text{-}MoS_2$ monolayer. We will show that the iCGD protocol can effectively address the errors of scan positions across the spectrum and help to achieve electron ptychography with high accuracy and precision.

# 1. Introduction

The ptychographic reconstruction of four-dimensional scanning transmission electron microscopy (4D-STEM) datasets has drawn great research attention because of its potential of getting wave-length limited spatial resolution[1–4], 3D imaging[2,5] and being dose-efficient[6–9]. In practice, achieving robust reconstruction and realizing quantitative data interpretation both require accurate knowledge of scan positions[10–12]. However, 4D-STEM experiments can be prone to scan positional errors. We previously reported a non-iterative method[13] that can effectively determine the global or uniform affine transformation between the scan-camera coordinates for atom-resolved 4D-STEM datasets. Iterative optimization of scan positions is still needed to achieve higher calibration accuracy and to address localized positional errors caused by issues such as non-uniform sample drifts and scan noise. We argue that existing iterative methods, such as simulated annealing[10], gradient descent (GD)[12,14], cross-correlation[15], and evolutionary algorithms[16], are not satisfactory for the cases of ptychographic reconstructions with atomic resolution. Systematic errors can be introduced when using these conventional methods due to the nature of the atom-resolved object, which contains strong features of atomic columns and a weak background (more details in *section 2.1*). In addition, iterative ptychographic reconstructions can be trapped into local minimum when various types of scan positional errors such as uniform affine transformations[17], non-uniform scan distortions[18–21], scan noise[22,23] and flags/skips[24,25] appear in the dataset. Last but not least, the existing iterative methods cannot effectively address the uniform/global affine transformation, especially when these methods are not supplied with a good enough first guess[13].

To overcome these problems, we hereby introduce an "integrated constrained gradient descent (iCGD)" protocol that consists of a series of CGD sub-routines[12,14], which were derived from the nature of 4D-STEM experiments and able to isolate and correct different types of positional errors. In this paper, we will first demonstrate the intrinsic limitations of the conventional GD method when applied to atom-resolved 4D-STEM datasets. This is followed by the sequential introduction and validation of different CGD sub-routines using a simulated in-focus 4D-STEM dataset with different types of pre-applied positional errors. Then the iCGD protocol integrating these CGD subroutines will be demonstrated using both simulated and experimental results from a 1H-MoS$_2$ monolayer. We will show that iCGD can effectively address the scan-positional errors across the spectrum and help to achieve high measurement accuracy and precision in electron ptychography.

## 2. Theory

*2.1 Introduction of the Conventional GD Method and Its Limitations*

Firstly, we will review how positional correction is realized in the conventional GD method and illustrate why systematic errors of scan positions occur. Let's assume that we have a thin object $o(r)$, the exit wave function $\psi_i(r)$ at the $i^{th}$ scan position can be approximated as the multiplication of the electron probe $p(r)$ and the object $o(r)$:

$$\psi_i(r) = o(r - r_i)p(r) \quad \quad EQ1$$

where $r$ is the position variable on the specimen plane, and $r_i$ represents the $i^{th}$ scan position. $o(r - r_i)$ is the shifted object function. Since the physical size of the object $o(r)$ is usually larger than that of the electron probe $p(r)$, the $o(r - r_i)$ is truncated to the same size as $p(r)$ during the numerical implementation of *EQ1*. In the electron microscope, the electron detector is placed in the far-field, so that the measured diffraction pattern $I_i(k)$ can be expressed as:

$$I_i(k) = \Psi_i(k)\Psi_i^*(k) \quad \quad EQ2$$

$\Psi_i(k)$ is the Fourier transform of exit wave $\psi_i(r)$, and $k$ is the reciprocal vector corresponding to $r$. Due to the positional error at the $i^{th}$ scan position, the diffraction pattern $I_i(k)$ estimated from the object $o(r)$ and probe $p(r)$ will deviate from the experimentally recorded diffraction pattern, $I_i^e(k)$. If the total number of diffraction patterns is *N*, correcting the $i^{th}$ scan position $r_i$ can be achieved by maximizing the log-likelihood $L(r_0, r_1..r_N, o, p)$ considering all *N* diffraction patterns[12,14]:

$$L(r_0, r_1..r_N, o, p) = -log(\prod_i(\sum(I_i(k) - I_i^e(k))^2))  \quad EQ3$$

In *EQ3*, the likelihood is simply expressed using the squared difference between the $I_i(k)$ and $I_i^e(k)$, and other types of likelihood functions can also be used to take other factors into account (*e.g.*, detector noise).[26,27] The direction for updating the scan position $r_i$ is chosen as the gradient of $L(r_0, r_1..r_N, o, p)$ with respect to the $i^{th}$ scan position:

$$\partial L/\partial r_i = (\partial L/\partial o(r-r_i))(\partial o(r-r_i)/\partial r_i) = G_i * o'(r-r_i) \quad EQ4$$

As shown in the above equation, the direction for updating $r_i$ is the multiplication of $G_i$, the gradient of the log-likelihood function with respect to the object function $o(r-r_i)$, and the spatial gradient of the object function $o'(r-r_i)$. $G_i$ is also the direction for updating the object function *o(r)* during iterative ptychographic reconstruction at position $r_i$. The step size used for positional correction along the update direction can be estimated by minimizing the modulus of another loss function $\mathcal{X}_i(r) = \psi_i(r) - \psi_{ic}(r)$, which is the difference between the estimated exit wave function $\psi_i(r)$ and modulus-constrained exit wave function $\psi_{ic}(r)$. Then, the shift-vector $dr_i$ can be acquired for correcting the $i^{th}$ scan position:

$$dr_i = \alpha \int p(r)o'(r-r_i)\mathcal{X}_i^*(r)/|p(r)o'(r-r_i)|^2 dr \quad EQ5$$

where α is the step size coefficient, and $o'(r)$ is the spatial gradient of the object. In practice, the area of the integration is kept the same as that of the electron probe *p(r)*. To compute $\mathcal{X}_i(r)$, the constrained exit wave function $\psi_{ic}(r)$ is generated by replacing the modulus of $\psi_i(r)$ with $I_i^e(k)$ in reciprocal space. $\mathcal{X}_i(r)$ is affected not only by the error in the scan positions but also by the remaining errors in the probe and object, which will affect the calculated shift vector $dr_i$. Since these factors are entangled with each other, it is very difficult to accurately optimize the probe, the object, and the scan positions all together with this conventional GD method. Although other methods use different forms of the updating functions for scan positions, [10,15,16] they still face similar challenges as the calculated shift vectors still heavily rely on local object features.

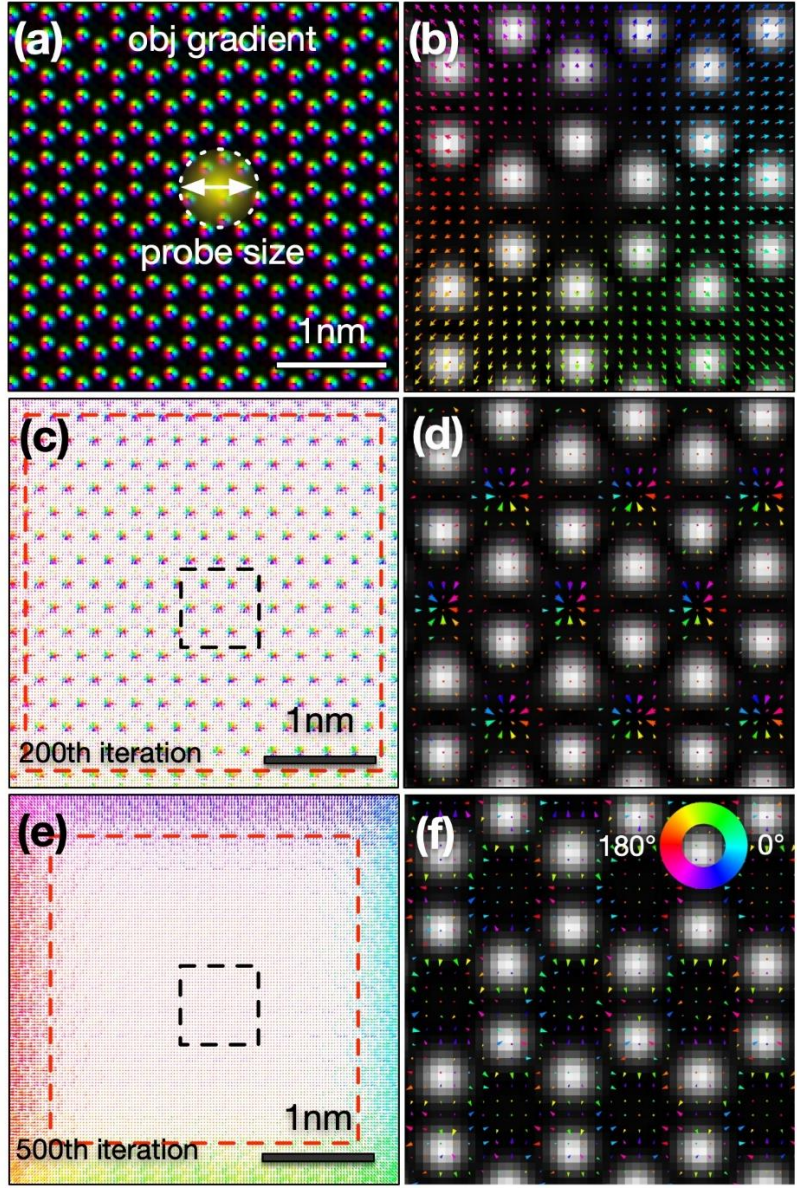

**Figure 1**. The inherent limitation of the conventional gradient descent (GD) method for the positional recovery, illustrated via a simulated atom-resolved 4D-STEM dataset of a 1H-MoS$_2$ monolayer. (**a**) The spatial gradient of the object function after 200 iterations of ePIE without enabling positional correction, overlaid with the probe size. (**b**) The zoomed view of the shift vectors added to scan positions (colored arrows) overlaid with object function. (**c, d**) The shift vectors calculated using *EQ5* at the 200$^{th}$ iteration. (**e, f**) The accumulated shift vectors calculated using *EQ5* after an additional 300 iterations of ePIE with the positional correction. (**d**) and (**f**) are zoomed views of the areas highlighted with black dashed lines in (**c**) and (**e**), respectively, overlaid with the same area of the object function as (**b**). The red dashed lines in (**c**) and (**e**) roughly highlight areas near the edge of the field of views that appear to better resemble the true shift vectors and have fewer periodic artifacts. The details for the 4D-STEM simulation are given in **Supporting Materials S1**.

Usually, the correction of scan positions is disabled at the start of the iterative reconstruction and is only carried out when the probe and object functions are stabilized after a certain number of iterations. However, applying this strategy alone does not solve the problem described above. As an example, a known distortion (in this case, 5% compression in both horizontal and vertical directions) was applied to scan positions of a simulated in-focus 4D-STEM dataset of a 1H-MoS$_2$ monolayer (**Figure 1**). This dataset was reconstructed using ePIE without enabling positional correction. The probe and object functions largely converged after 200 iterations and the resulting spatial gradient of the object ($o'(r)$ in *EQ4* and *EQ5*) is shown in **Figure 1(a)**. **Figure 1(b)** shows a zoomed view of the object function overlaid with the true shift vectors. The correction of scan positions was then switched on and another convergence is reached after at about the 500$^{th}$ iteration (**Supporting Materials S2**). The shift vectors calculated using *EQ5* at the 200$^{th}$ and the 500$^{th}$ iteration are shown in **Figures 1(c, d)** and (**e, f**), respectively. The shift vectors in both cases clearly deviate from their pre-applied values and show strong correlations with the object function, which can be better viewed in the zoomed views shown in **Figures 1** (**d**) and (**f**). This "correlation effect" originates from *EQ4* and *EQ5*. Interestingly, at the area near the edges (highlighted with red-dash lines in **Figures 1(c)** and (**e**)), the shift vectors (*i.e.*, their color-coded directions) appear to better resemble the true distortion values compared to the central areas. Although the size of these "good areas" has increased slightly at the 500$^{th}$ iteration compared to that at the 200$^{th}$ iteration, the "correlation effect" still dominates and will introduce artifacts in the final reconstruction results, such as arbitrary resolution extensions or even reconstruction failures (**Supporting Materials S3**). As a result, the conventional GD method is not satisfactory in correcting positional errors for atom-resolved electron ptychography and a better method is needed.

*2.2 Introduction of Constrained Gradient Descent (CGD) methods.*

Since most of the 4D-STEM experiments, if not all, use a raster scanning pattern, we take advantage of *a priori* knowledge of the nature of such an experiment to correct errors of scan positions. **Figure 2** summarizes typical types of positional errors in STEM and existing treatment methods for STEM experiments, including the uniform affine transformation, non-uniform scan distortions (low-frequency positional errors) and scan noise (high-frequency positional errors). A notable type of positional errors is called flags and skips[18,19,24,25], which is a sudden shift of the row caused by the hysteresis and alignment error of the scanning system, along the horizontal and vertical directions, respectively. In the following sections, we will introduce different constraints that can be applied to the shift vectors computed using the conventional GD method in order to isolate and correct different types of positional errors. These constraints include the *affine constraint*, the *low-pass-filter (LPF) constraint*, the *line constraint* and the *high-pass-filter (HPF) constraint*. Subsequently, we will introduce a suggested workflow of the iCGD protocol that integrates these constraints. We will show that the iCGD protocol can effectively eliminate complex mixtures of various positional errors in experiments.

| | Low frequency | | | High frequency |
|---|---|---|---|---|
| positional error types | Uniform affine transformation | Non-uniform scan distortion (low-frequency positional error) | Flags and skips | Scan noise (high-frequency positional error) |
| methods for conventional STEM | Multiple frame based method (Sang et.al)[17] | Multiple frame based non-rigid registration (B. Berkels, et.al, L. Jones, et.al.) [19, 20, 21] | Single frame. (N. Braidy et.al) [24] Orthogonal pair. (C. Ophus et.al) [25] | Single frame method: Jitterbug. (L. Jones, et.al) [22] |
| CGD methods | affine constraint | LPF constraint | line constraint | HPF constraint |

**Figure 2**. The overview of different types of positional errors in STEM-related experiments and their corresponding correction methods both in literature and in this work.

*2.2.1 The Affine Constraint for the Correction of Uniform Affine Transformations*

We previously reported a non-iterative method to correct the uniform affine transformation in scan positions[13]. As we will show later, the quality of the ptychographic reconstruction will still be limited by residual affine transformations, which need to be removed using an iterative method. We hereby introduce the *affine constraint* that can isolate the uniform affine transformation component from the shift vectors obtained from the conventional GD (***EQ5***). As shown in ***EQ6***, the uniform affine transformation can be defined with a 3 by 2 matrix $A$. The $(x_i, y_i)$ and the $(dx_i, dy_i)$ are the components of the scan position $\boldsymbol{r}_i$ and its corresponding shift-vector $d\boldsymbol{r}_i$ along the horizontal and vertical directions of pixelated detectors, respectively.

$$\begin{bmatrix} dx_i \\ dy_i \end{bmatrix} = \begin{bmatrix} x_i \\ y_i \\ 1 \end{bmatrix} A \quad \boldsymbol{EQ6}$$

To apply the *affine constraint*, the 4D-STEM dataset will be first processed using the conventional GD method so that the unconstrained shift vector $d\boldsymbol{r}_i$ is calculated using ***EQ5***. Then the affine transformation component (*i.e.,* the matrix $A$) of shift vectors can be obtained using ***EQ6*** via least square fitting (for more details see **Supporting Materials S4**). After that, the constrained shift vector $d\boldsymbol{r}_{ic}$ can be calculated using ***EQ7***. ***EQ7*** appears very similar to ***EQ6*** except for an additional weight factor which contains a weight factor $W_A$. The purpose of ***EQ7*** is to calculate the constrained shift vectors, while the purpose of ***EQ6*** is to use pre-determined shift vectors to find the affine transformation component (*i.e.*, the matrix $A$).

$$\begin{bmatrix} dx_{ic} \\ dy_{ic} \end{bmatrix} = W_A \begin{bmatrix} x_i \\ y_i \\ 1 \end{bmatrix} A \quad \boldsymbol{EQ7}$$

To showcase the effectiveness of the *affine constraint*, we tested 3 different types of affine transformations added to scan positions in the simulated 4D-STEM dataset used in **Figure 1**. The unconstrained shift vectors calculated using ***EQ6*** at the 200$^{\text{th}}$ iteration are shown for the cases where scan positions are compressed (**Figure 3 (a)**), sheared (**Figure 3 (b)**), or rotated (**Figure 3 (c)**), respectively. The "correlation effect" can be clearly observed, similar to the case of **Figure 1**. As shown in **Figures 3 (d)**, **(e)**, and **(f)**, respectively, the shift vectors with the *affine constraint* resemble much more closely to their corresponding ground truths. These datasets were further processed using ePIE within a few hundred iterations, with the positional correction enabled. As shown in **Supporting Materials S5** and **S6**, applying the *affine constraint* gives better reconstructions with smaller residual errors.

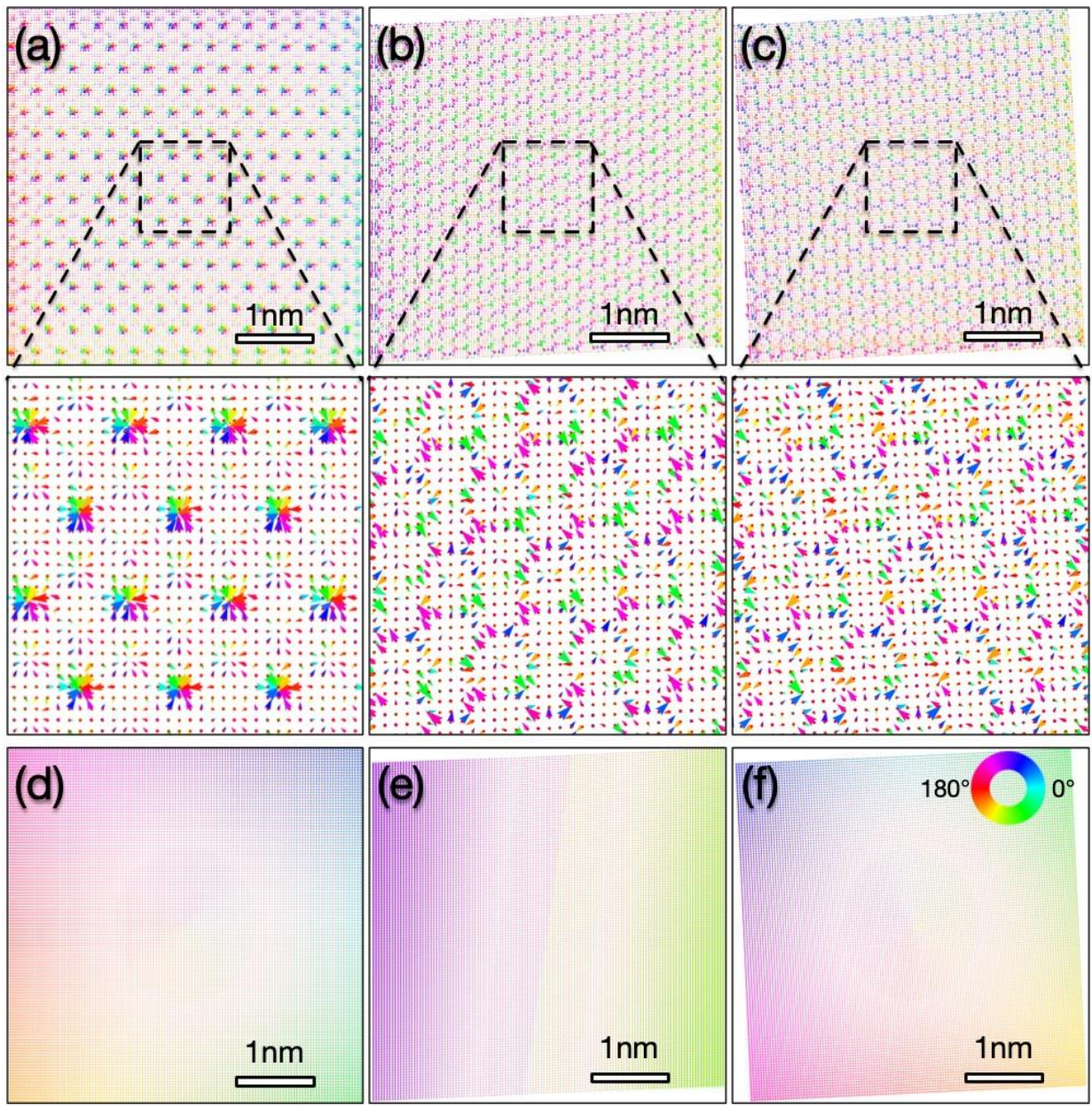

**Figure 3**. A comparison between the unconstrained and affine-constrained shift vectors for the in-focus 4D-STEM dataset of the 1H-MoS$_2$ monolayer. (**a-c**) are snapshots of the shift vectors at the 200$^{th}$ iteration when uniform compression, shearing, and rotation are applied to the initial scan positions, respectively. The obvious periodicity of MoS$_2$ is observed in the zoomed view of shift

vectors within the black rectangles in (**a-c**). The corresponding affine-constrained shift vectors are shown in (**d**), (**e**), and (**f**), respectively.

It is important to note that we cannot just apply the affine-constrained shifting vectors to scan positions, otherwise the iterative ptychographic reconstruction can still shift the scan positions back, giving little or no improvement in the end. Following the idea of classical methods such as the serial cross-correlation method [15] and the annealing algorithm [10], the object function $o(r)$ will also need to be deformed in the same way as scan positions in each iteration, as shown in $EQ8$:

$$o'(r_i + dr_{ic}) = |o(r_i)| \; EQ8$$

$o(r)$ is the updated object function after all scan positions have been processed in each iteration, and $o'(r)$ is the deformed object function. Only after applying shift-vectors on both scan positions and the object function, significant improvement in the iterative reconstructions can be seen, as shown in **Figure S6** of the **Supporting Materials S5**.

The weight factor $W_A$ defined in $EQ7$ can help to balance the speed and the quality of the convergence for the iterative reconstruction. In practice, we get the best results when the $W_A$ is set to be larger than 1. However, if the $W_A$ is set too large, the algorithms may not be able to find the correct scan positions and the reconstruction becomes unstable. The optimum value of $W_A$ is likely to vary from case to case. More discussion about this can be found in **Supporting Materials S6**.

*2.2.2 The Line Constraint for the Correction of Skips and Flags*

Next, we will introduce the *line constraint* that is designed to address the "flags" (**Figure 4** (**a**)) and "skips" (**Figure 4** (**b**)), which can be considered as misplacing the entire row of pixels due to alignment errors or hysteresis in the scan coils[18,19,24,25]. To implement this, we average the shift vectors from pixels in the same row and then reassign them with these averaged values to force these pixels in the same row to shift together. The line-constrained shift vector $dr_c$ for pixels within a particular row can be formulated as follows:

$$d\boldsymbol{r}_c = \begin{bmatrix} dx_{ic} \\ dy_{ic} \end{bmatrix} = W_L \sum_{i=1}^{L} d\boldsymbol{r}_i / L \quad \boldsymbol{EQ9}$$

In *EQ9*, $L$ is the number of pixels in the fast-scanning direction, $W_L$ is the weight factor that can be tuned to achieve both good results and reasonable convergence rates, just like $W_A$ discussed in the previous section. To show the effectiveness of the *line constraint*, we will use the same simulated in-focus 4D-STEM dataset of the 1H-MoS$_2$ monolayer discussed in **Figure 1**, except that random flags and skips were added to scan positions. The unconstrained shift vectors will be obtained from *EQ5* after the dataset was first reconstructed using ePIE without enabling positional correction. We can then shift all the pixels together in the same scanning row using the averaged shift vectors obtained from *EQ9*. Following the idea of L. Jones *et al.*[19], we found that better results can be obtained by allowing the pixels in the same row to be shifted in a locally correlated way instead of applying the same shift to all pixels. To do that, we can divide the row into a certain number of pieces with equal line widths (**Figure 4 (c)**). An averaged shift vector can be obtained in each piece and then the line-constrained shift vector for each pixel will then be obtained using interpolation based on these averaged values. *EQ9* can be considered as the special case where the number of pieces is set to 1, meaning that the line width equals the total number of pixels in the row.

**Figure 4.** (**d, e**) and **Figure 4.** (**f, g**) show the shift vectors before and after applying the *line constraint*, for two cases of having random flags and skips added to scan positions in the simulated 4D-STEM data. The *line constraint* can effectively remove the periodic artifacts (*i.e.*, the correlation effect) that can be seen in **Figures 4(d) and 4(f)**. As shown in **Supporting Materials S7**, with the *line constraint*, the iterative construction can achieve faster convergence with smaller residual errors, which can help to avoid artifacts in the result. When implementing the *line constraint*, the object is not deformed along with scan positions in each iteration, which is different from the *affine constraint*.

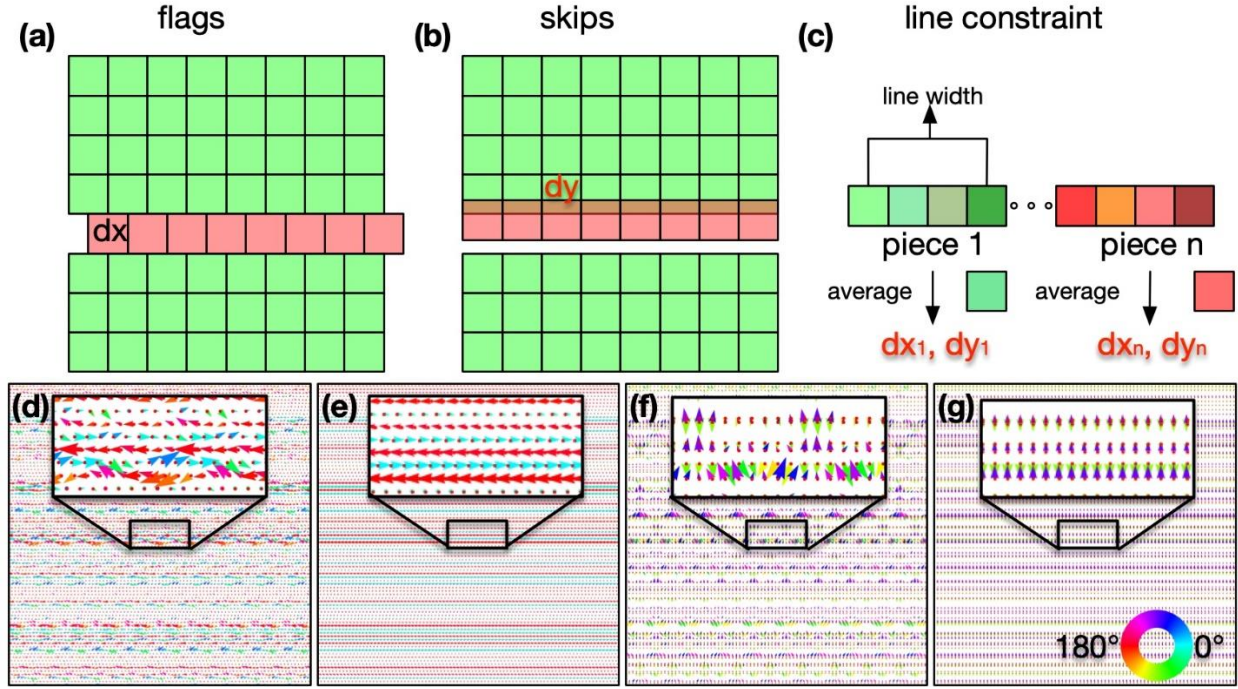

**Figure 4**. The *line constraint* designed to correct flags (**a**) and skips (**b**) components of positional errors. (**c**) The illustration of the strategy to break the row into a certain number of pieces with the same line width. The line-constrained shift vectors for each pixel within the row will be obtained via interpolation. (**d**) and (**f**) are the computed distribution of shift vectors at the 200$^{th}$ iteration using the GD method when random flags and skips are applied to scan positions, respectively. (**e**) and (**g**) illustrate the shift vectors after applying the *line constraint* with the number of pieces set to 1 for each row for the cases of (**d**) and (**f**), respectively.

*2.2.3 Low-Pass Filter Constraint (LPF) for Correcting non-uniform scan distortions.*

The *low-pass filter constraint* (LPF) is proposed to remove non-uniform distortions of scan positions, which are in the low-frequency regime of positional errors (**Figure 2**). The LPF basically extracts the low-frequency components of shift vectors computed using the GD method in each iteration. In the implementation of the *LPF constraint*, two 2D arrays densely sampled on the grid mesh of the object function are first generated for both the horizontal and vertical components of shift vectors. Then the low-pass-filtering is applied to these two arrays with a maximum spatial frequency $k_l$ lower than that of the Bragg peaks nearest to the center (**Figure 5** (**a**)). Consequently, the correlation effect is avoided by eliminating atomic features of shift vectors in both horizontal and vertical directions. Finally, the two components of shift vectors with the *LPF constraint* are right the values of these two low-pass-filtered arrays at each scan position. When implementing the *LPF constraint*, the object also needs to be deformed with scan positions in each iteration, just like the case of the *affine constraint*.

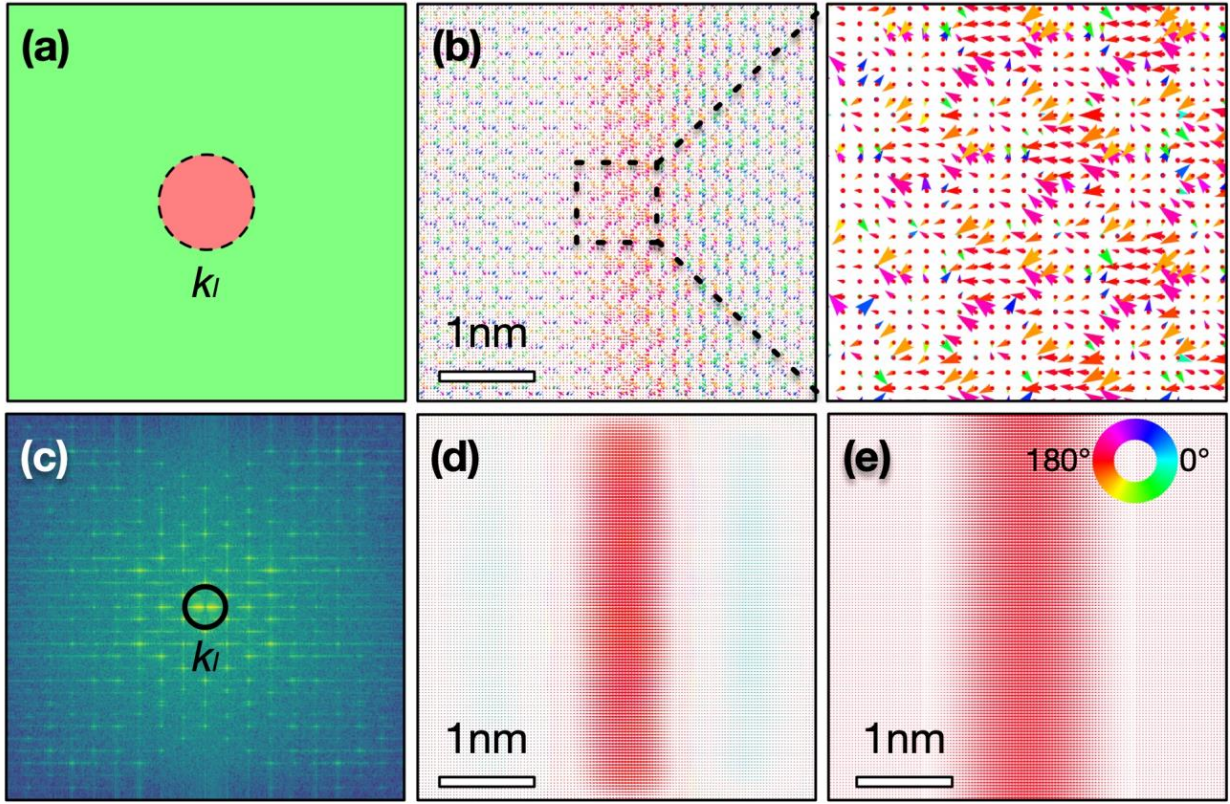

**Figure 5**. Demonstration of the *LPF constraint* using the simulated 4D-STEM dataset with a Gaussian-shaped displacement field along the horizontal direction applied to the center part of the scan positions. (**a**) In the *LPF constraint*, the low-frequency components of shift vectors are retained, and the information related to the atom lattice is eliminated when the maximum frequency is set to $k_l$. (**b**) and (**c**) are the shift vectors computed by the GD method and the summed power spectrum of their corresponding 2D arrays of shift vectors along both the horizontal and vertical directions. Bragg peaks can clearly be seen in (**c**) due to the inherent correlation effect of the GD

method. After applying the LPF, the smooth distribution of shift vectors is recovered in (**d**). After 500 iterations, the pre-applied displacement is fully recovered as shown in (**e**).

To illustrate the effectiveness of the *LPF constraint*, a Gaussian-shaped displacement field distributed along the horizontal direction is added to scan positions of the simulated 4D-STEM dataset. The sigma value of the Gaussian-shaped displacement field is set to 7 Å, its center matches the horizontal center of the scanned area, and the amplitude is set to 0.5 Å. Without applying any constraints, the shift vectors computed using the GD method at the $200^{th}$ iteration of ePIE are plotted in **Figure 5**(**b**). Just like the cases in *sections 2.2.1* and *2.2.2*, an arbitrary periodicity associated with the lattice structure of $MoS_2$ can be seen. After applying the LPF (**Figure 5**(**c**)), these artifacts in the displacement field are removed and a smoothly distributed displacement field is recovered, as shown in **Figure 5**(**d**). Additional iterations are needed to remove the artifacts caused by the filtering in Fourier space. The pre-applied non-uniform deformation was fully identified after 500 iterations, as shown in **Figure 5**(**e**).

*2.2.4 High-Pass Filter (HPF) Constraint for the Correction of Scan Noise.*
Finally, we propose a *HPF constraint* to extract and eliminate the high-frequency components correlated with the reconstructed object. As was the case in the section on the *LPF constraint*, two 2D arrays will first be obtained for both the horizontal and vertical components of shift vectors via spline interpolation. While the *LPF constraint* has handled the positional errors with spatial frequencies less than $k_l$, the *HPF constraint* will isolate and allow the correction of the positional errors with higher spatial frequencies. The shift vectors with the *HPF constraint* can also be obtained via an inverse Fourier transformation and a conversion from the gridded data to scattered values at corresponding scan positions. Different from the previously mentioned *affine*, *line* and *LPF constraints*, the *HPF constraint* is applied to the shift vectors only once, usually at the last stage of the reconstruction. Similar to the *line constraint*, the object is not deformed using *EQ8* along with scan positions in the *HPF constraint*.

To demonstrate the effectiveness of the *HPF constraint*, random perturbations were added to the scan positions of the simulated in-focus 4D-STEM dataset used in previous sections. After the initial reconstruction of the dataset without positional correction, the unconstrained shift vectors calculated via *EQ5* are shown in **Figure 6(a)** and the summed power spectrum of their corresponding horizontal and vertical 2D arrays is plotted in **Figure 6(b)**. The key step for HPF is to remove the "Bragg peaks" caused by the "correlation effect". To do that, the radial average of the power spectrum was first obtained. At a specific radial distance from the center, pixels with intensities larger than a certain threshold above the radial average will be set to a new intensity value. This value equals the newly obtained radial average from the remaining pixels at that radial distance. As shown in **Figure 6(c)**, the Bragg peaks were successfully removed. Compared to the genuine solutions of the random distortions (**Figure 6(d)**), the *HPF-constrained* shift vectors (**Figure 6(e)**) show no periodic artifacts and largely resemble the genuine solution, which is demonstrated in their difference map that is close to 0 (**Figure 6(f)**).

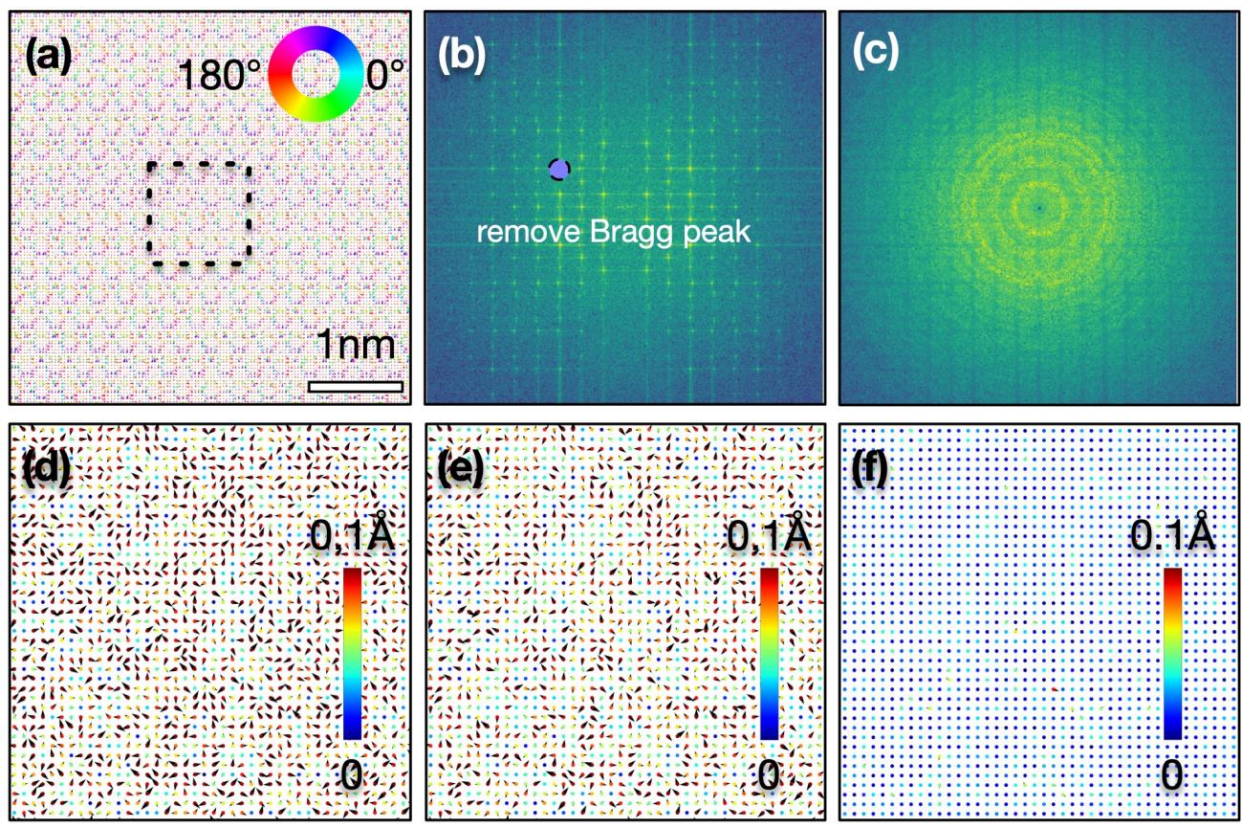

**Figure 6**. The *HPF constraint* designed to isolate and correct for high-frequency positional errors (scan noise). (**a**) The computed distribution of shift vectors via the GD method for the in-focus 4D-STEM dataset of the 1H-MoS$_2$ monolayer with random scan noise added to scan positions. The applied positional errors follow the random linear distributions, and the amplitude is 0.1Å. (**b**) and (**c**) are the summed power spectrum of the horizontal and vertical shift 2D arrays before and after applying the *HPF constraint*, respectively. (**d**) shows the genuine value of the random shift vectors applied to the middle region marked by the black, stripped rectangle in (**a**). (**e**) is the map of *HPF-constrained* shift vectors. The difference between (**d**) and (**e**) is plotted in (**f**). In (**d**), (**e**) and (**f**), both the color and the lengths of the shift vectors indicate the amplitude of displacement.

*2.3 The suggested workflow of the iCGD protocol to treat 4D-STEM ptychography datasets in general.*

In the sections above, different CGD sub-routines to treat different types of positional errors were introduced. These routines need to be incorporated into the ptychographic reconstruction workflow to process experimental 4D-STEM datasets, which can have a mixture of different types of positional errors across the frequency spectrum (**Figure 2**). A suggested workflow for iCGD is shown in **Figure 7**. The 4D-STEM dataset needs to be processed using the non-iterative method we reported earlier[13], so that the iteration method can be better initialized with the majority of positional errors due to affine transformations eliminated. The 4D-STEM dataset will then undergo iterative reconstruction with positional corrections using the iCGD protocol, which contains four stages. The details of these four stages are schematically shown in **Figure 7(b)**. At the first stage, the dataset will be iteratively reconstructed without shifting scan positions. This will allow the initial functions of the object and the probe to be obtained. From that, the calculation of the unconstrained shift vectors using *EQ5* is conducted in each iteration of the second stage. At each iteration, the *affine*, *line* and *LPF constraints* will be applied to the unconstrained shift vectors to obtain three corresponding sets of constrained shift-vectors. The scan positions will be updated with all three sets of constrained shift vectors, while the object will be deformed according to the combined shift vectors with *affine* and *LPF constraints*. In the third stage of the reconstruction, the iterative reconstruction with the correction of scan positions is first carried out with the conventional GD method based on *EQ5* between B and C. At the $C^{th}$ iteration, the *HPF constraint* is applied to the accumulated shift vectors during the B to C iteration. The final stage of the reconstruction will start from the state of the $B^{th}$ iteration, and the corresponding scan positions will be updated for the last time using the *HPF-constrained* shift vectors at the $C^{th}$ iteration. Lastly, in stage four, the probe and object will be optimized until convergence. We practically found that having the iterations between B and C can help the *HPF constraint* to better identify and correct high-frequency positional errors. The effectiveness of this suggested iCGD workflow is validated on the simulated 4D-STEM dataset, and the pre-applied mixture of positional errors is well recovered as shown in **Supporting Materials S8**.

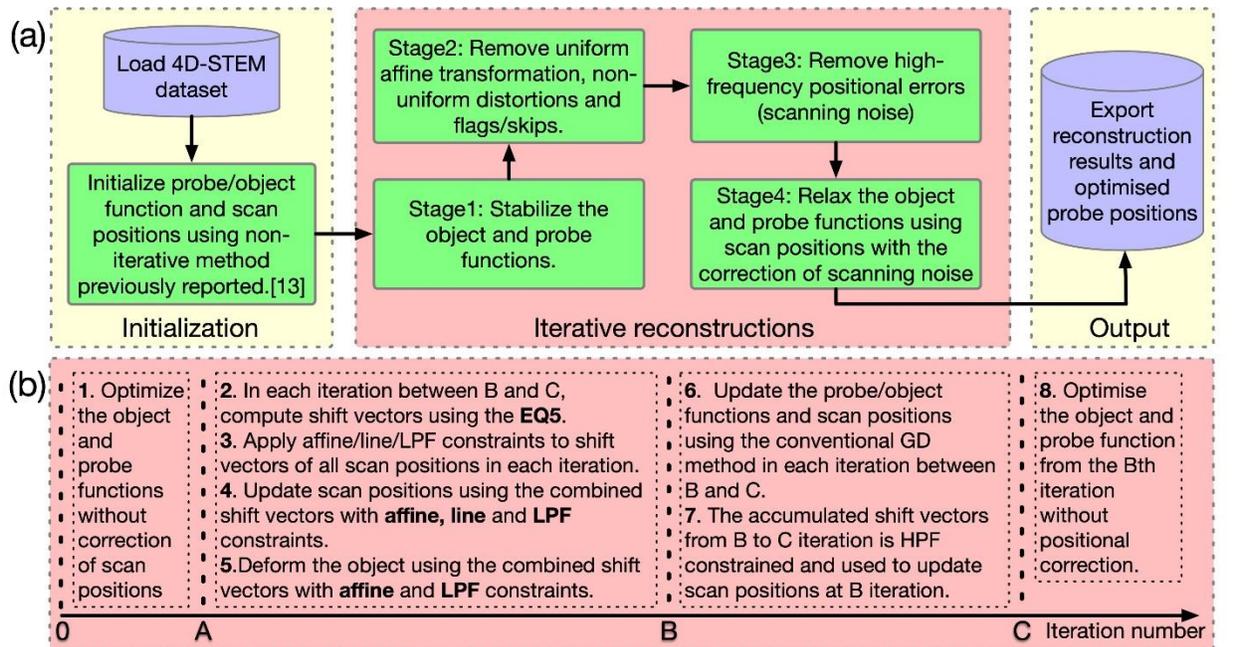

**Figure 7**. (**a**) The overall workflow for the iterative ptychographic reconstruction with positional correction using the iCGD protocol. The 4D-STEM dataset will first undergo the initialization using the previously reported non-iterative method (geometric calibration [13]) to largely remove the uniform affine transformations of scan positions. (**b**) The suggested workflow for the iterative reconstruction with iCGD protocol in more detail. The labels A, B, C indicate the iteration numbers that may vary from case to case.

## 3. Application of the iCGD protocol on a practical 4D-STEM dataset.

The iCGD protocol was applied to an experimental in-focus 4D-STEM dataset of the MoS$_2$ monolayer, obtained using Merlin Medipix3 1R detector installed on an aberration-corrected JEOL ARM200CF microscope operated at 80kV. More experimental details about this dataset can be found in **Supporting Materials S9**. The dataset was processed according to the workflow shown in **Figure 7**. In the iterative reconstruction, the first stage was carried out from the 0$^{th}$ to the 50$^{th}$ iteration using ePIE without the correction of scan positions. The initial unconstrained set of shift vectors was obtained to start the second stage of the iteration, which covers the 50$^{th}$ to the 1850$^{th}$ iterations. During this stage, the *line*-, the *LPF* and the *affine constraints* were applied, and the combined constrained shift vectors were used to update scan positions at each iteration. The object is simultaneously deformed according to the combined shift vectors with *LPF* and *affine constraint*s. For the implementation of the *line constraint*, the number of pieces will be set to 1 and will increase by 1 per 400 iterations. The weight factor $W_L$ was set to 0.5. For the *affine constraint*, the $W_A$ was set to 4.0. The third stage of the reconstruction lasted from the 1850$^{th}$ to the 2150$^{th}$ iteration. The *HPF constraint* was applied once to the collective shift vectors (i.e. the difference map between the scan positions of the 2150$^{th}$ and the 1850$^{th}$ iteration states). The final stage of the reconstruction started from the results at the 1850$^{th}$ iteration, after updating the scan positions for the final time using the *HPF-constrained* shift vectors. The functions of the probe and object will be optimized until convergence. For this dataset, the final stage lasts for 100 iterations.

The reconstruction results are shown in **Figure 8** and **Supporting Materials S10**. The averaged phase distributions of the 1H-MoS$_2$ lattice in retrieved objects using the conventional GD method and the iCGD protocol are plotted in the bottom left corners of **Figure 8(c)** and **(d),** respectively. In the case where the iCGD protocol was adopted, the 6-fold symmetry of the 1H MoS$_2$ lattice is well recovered and a central symmetric phase distribution of atom columns is observed. The accuracy and precision of the results were evaluated by measuring the bond distances and bond angles of Mo's three nearest S neighbors, which was obtained from fitting Gaussian functions to the reconstructed phase image. As shown in **Figure 8(a)**, the phase image obtained from the reconstruction with the iCGD protocol gave a much narrower distribution of the Mo-S bond distance (standard deviation 1.9pm), compared to the counterparts reconstructed with the conventional GD method (standard deviation 5.4pm). Similarly, when measuring the projected bond angle of Mo-S, the data processed by the iCGD protocol also show a much narrower distribution the correct value of 120°, compared to the counterpart processed by the conventional GD method (**Figure 8(b)**.

The measurement accuracy of the result via the iCGD protocol can also be visualized using the difference map between the reconstructed result from an ideal lattice template. To do that, the averaged basis vectors are first obtained from the reconstructed phase images, and then an ideal lattice template is generated from a reference point. Then the displacement vectors of each unit cell from the ideal lattice template can be plotted, which are shown in **Figures 8(c)** and **(d)**, for the cases where the GD and iCGD protocols were adopted, respectively. In these two plots the reference points were at the identical location and were marked by red circles. The displacement vectors in the two plots share the same scale. From these results, we conclude that electron ptychography high accuracy and precision is experimentally achieved with the iCGD protocol.

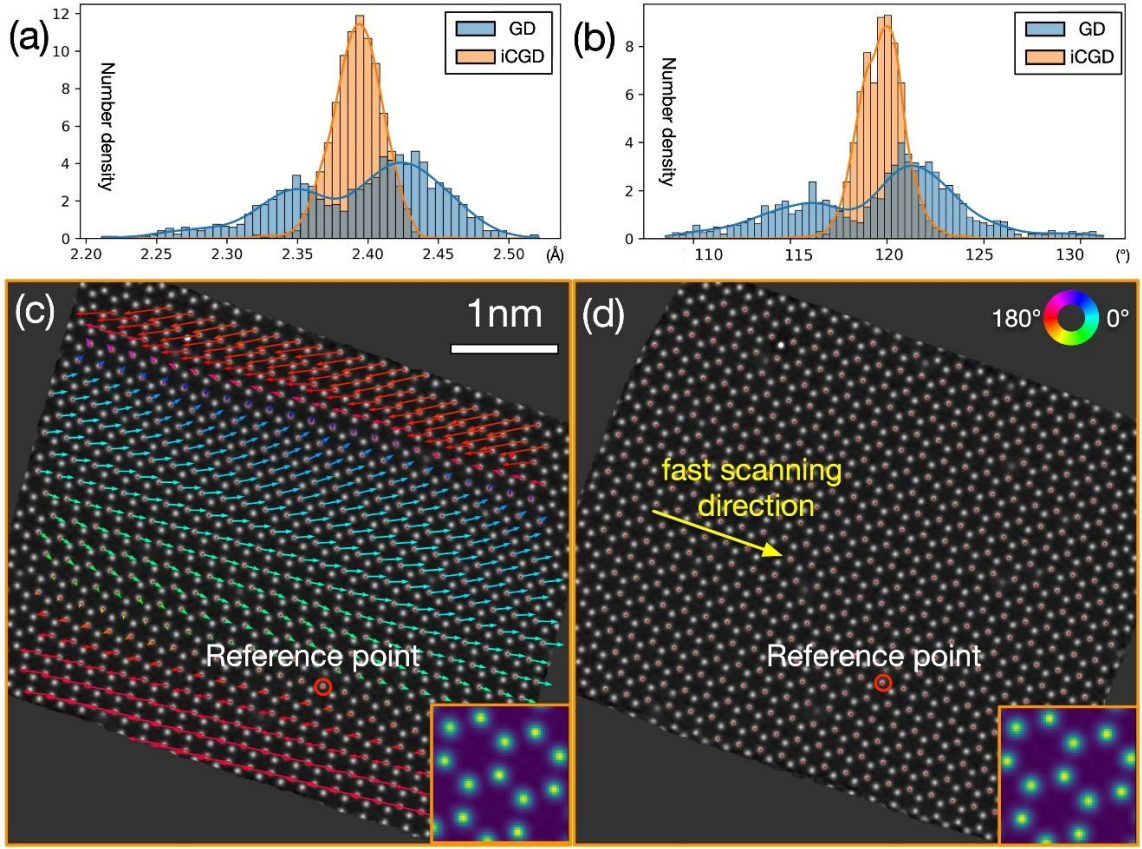

**Figure 8**. Experimental demonstration of high-precision electron ptychography achieved by the iCGD protocol. (**a**) and (**b**) are the histograms of measured Mo-S distances and angles from the reconstructed phase images of the 1H-MoS$_2$ monolayer. The standard deviations of the Mo-S distances on the reconstructed phase using our iCGD protocol and conventional GD method are 1.9 pm and 5.4 pm, respectively. The maps of the displacement vectors calculated from the ideal lattice template and the measured phase image for the cases constructed with (**c**) the conventional

GD method and **(d)** the iCGD protocol, respectively. The averaged phase images of the 1H-MoS$_2$ monolayer for (**c**) and (**d**) are shown as the insets.

## 4. Summary

In this paper, we have introduced an iCGD protocol with several CGD sub-routines covering various positional errors in order to achieve high accuracy and precision in electron ptychography,. While the conventional GD method is shown to produce systematic errors when dealing with atom-resolved 4D-STEM datasets, the iCGD protocol can overcome such limitations and eliminate complex scan-positional errors across the spectrum during the iterative electron ptychography reconstruction. Higher measurement precision and fewer artifacts were experimentally demonstrated in the 1H-MoS$_2$ monolayer using the iCGD protocol compared to the case where the conventional GD method was used.

The iCGD protocol is currently only applied to 4D-STEM experiments where a raster scanning pattern is used. No *a priori* knowledge about the structure of the material was needed for the iCGD except that the sample is assumed to be crystalline in *the HPF constraint*. Like many other positional correction methods, the iCGD protocol can also fail if the starting scan positional errors are too large (**Supporting Materials S11**). Nevertheless, we believe the iCGD protocol is a solid development toward robust electron ptychography that can provide high measurement accuracy and precision by eliminating scan positional errors and can reduce reconstruction artifacts.

**Acknowledgment:**
Q. He would also like to acknowledge the support of the National Research Foundation (NRF) Singapore, under its NRF Fellowship (NRF-NRFF11-2019-0002). F. Zhang acknowledges the support of the National Natural Science Foundation of China (11775105, 12074167). M.B. acknowledges support from the Singapore Ministry of Education Academic Research Fund (Tier 1 startup project R-284-000-179-133). The authors thank Stephen John Pennycook for the valuable discussions on the algorithm, and Matus Krajnak of Quantum Detectors Ltd for guidance and setup of the MerlinEM detector used on the JEOL ARM200CF.